\documentclass[11pt,a4paper]{article}
\pdfoutput=1
\usepackage{jcapmod}

\usepackage{booktabs}
\usepackage[english]{babel}
\usepackage{amsmath,amssymb,amsbsy,amstext, amsthm, simplewick}
\usepackage{hyperref}
\usepackage{graphicx}
\usepackage{amsfonts}
\usepackage{amssymb}
\usepackage[small]{caption}
\usepackage{upgreek}
\usepackage[svgnames,dvipsnames,x11names,table]{xcolor}
\usepackage{multirow}
\usepackage{geometry}
\usepackage[hang,flushmargin]{footmisc}
\usepackage{comment}
\usepackage{braket}
\usepackage{pifont}
\usepackage{amssymb}
\usepackage{dirtytalk}
\usepackage{authblk}

\usepackage{float}

\def\Fc{\mathcal{F}}

\def\Kc{\mathcal{K}}
\def\Lc{\mathcal{L}}

\def\Fc{\mathcal{F}}

\def\Kc{\mathcal{K}}
\def\Lc{\mathcal{L}}

\newcommand{\cmark}{\ding{51}}%
\newcommand{\xmark}{\ding{55}}%

\newcommand{\ba}{\begin{aligned}}
\newcommand{\ea}{\end{aligned}}

\newcommand{\cF}{{\cal F}}

\newcommand{\be}{\begin{equation}}
\newcommand{\ee}{\end{equation}}

\newcommand{\bea}{\begin{eqnarray}}
\newcommand{\eea}{\end{eqnarray}}

\usepackage{colortbl}
\definecolor{lightgreen}{cmyk}{0.2, 0, 0.2, 0.2}
\definecolor{lightgray}{cmyk}{0.1,0.2,0,0.1}
\definecolor{lightgray2}{cmyk}{0.1,0.1,0,0.1}

\makeatletter
\newlength{\apb@width}
\newcommand{\autoparbox}[2][c]{\settowidth{\apb@width}{#2}\parbox[#1]{\apb@width}{#2}}

\makeatother

\definecolor{lightgray}{gray}{0.9}

\usepackage[framemethod=default]{mdframed}
\newmdenv[skipabove=7pt,
skipbelow=7pt,
rightline=false,
leftline=false,
topline=false,
bottomline=false,
backgroundcolor=gray!10,
linecolor=gray,
innerleftmargin=5pt,
innerrightmargin=5pt,
innertopmargin=5pt,
innerbottommargin=5pt,
leftmargin=0cm,
rightmargin=0cm,
linewidth=4pt]{eBox}

\numberwithin{equation}{section}

\def\beq{\begin{equation}}
\def\eeq{\end{equation}}

\def\bea{\begin{eqnarray}}
\def\eea{\end{eqnarray}}

\newcommand{\pd}{\partial}

\def\beq{\begin{equation}}
\def\eeq{\end{equation}}
\def\bea{\begin{eqnarray}}
\def\eea{\end{eqnarray}}

\DeclareRobustCommand{\SkipTocEntry}[4]{}

\definecolor{blue3}{RGB}{31, 119, 180}
\definecolor{red3}{RGB}{	214, 39, 40}
\definecolor{orange3}{RGB}{255, 127, 14}
\definecolor{green3}{RGB}{44, 160, 44}

\begin{document}

\begin{titlepage}
\setcounter{page}{1} \baselineskip=15.5pt
\thispagestyle{empty}

\begin{center}
{\fontsize{18}{18} \bf An anisotropic bouncing universe in non-local gravity}
\end{center}

\vskip 20pt
\begin{center}
\noindent
{\fontsize{12}{18}\selectfont K. Sravan Kumar$^{1,2}$, Shubham Maheshwari$^{2}$, Anupam Mazumdar$^{2}$ and Jun Peng$^{2}$}
\end{center}

\begin{center}
 \vskip 8pt
  \textit{$^{1}$Department of Physics, Tokyo Institute of Technology, 2-12-1 Ookayama, Meguro-ku, Tokyo 152-8551, Japan} \\
\textit{$^{2}$Van Swinderen Institute for Particle Physics and Gravity, University of Groningen, Nijenborgh 4, 9747 AG Groningen, The Netherlands}\\
\textit{Email: sravan.k.aa@m.titech.ac.jp, s.maheshwari@rug.nl, anupam.mazumdar@rug.nl, jun.peng@rug.nl}
\end{center}

\vspace{0.4cm}
\begin{center}
\today

{\bf Abstract}
\end{center}

\noindent

We show that it is possible to realize a cosmological bouncing solution in an anisotropic but homogeneous Bianchi-I background in a class of non-local, infinite derivative theories of gravity. We show that the anisotropic shear grows slower than in general relativity during the contraction phase, peaks to a finite value at the bounce point, and then decreases as the universe asymptotes towards isotropy and homogeneity, and ultimately to de Sitter. Along with a cosmological constant, the matter sector required to drive such a bounce is found to consist of three components - radiation, stiff matter and $k$-matter (whose energy density decays like the inverse square of the average scale factor). Generically, $k$-matter exerts anisotropic pressures. We will test the bouncing solution in local and non-local gravity and show that in the latter case it is possible to simultaneously satisfy positivity of energy density and, at least in the late time de Sitter phase, avoid the introduction of propagating ghost/tachyonic modes.

\end{titlepage}

\restoregeometry

\setcounter{tocdepth}{2}

\section{Introduction}

In cosmology, the inevitable existence of the Big Bang singularity in the presence of standard matter fluids is a fundamental problem in General Relativity (GR) \cite{Hawking:1969sw}. Even in vacuum, GR has Kasner type solutions which lead to the (in)famous Belinsky-Khalatnikov-Lifshitz (BKL) singularity \cite{Kasner:1921zz}.
Several modifications of GR with higher curvature terms or hypothetical matter are very successful in explaining the cosmic microwave background and large scale structure of the universe by invoking a phase of primordial inflation \cite{Aghanim:2018eyx} (see~\cite{Mazumdar:2010sa} for a review). However, the problem of an initial singularity still persists since the standard inflationary epoch is geodesically past incomplete \cite{Borde:2001nh,Muller:2017nxg}. One of the assumptions of the inflationary paradigm is that it requires the initial existence of a sufficiently homogeneous and isotropic patch of spacetime described by a Friedmann-Lema\^itre-Robertson-Walker (FLRW) metric. This is the problem of initial conditions \cite{Vachaspati:1998dy}. In fact, according to Misner's initial chaos, there is no a priori reason to assume any spacetime symmetries (such as FLRW) for the universe to begin with \cite{Misner:1970yt,Misner:1969hg}. The problem of cosmological singularity can be addressed in quantum cosmology through the no boundary proposal \cite{Hawking:2000bb,Hartle:1983ai}. Another interesting way of resolving the cosmological singularity is through a non-singular bouncing scenario \cite{Battefeld:2014uga} which has been widely investigated in several ultraviolet (UV) completions of GR such as string theory \cite{Gasperini:2002bn}, string gas cosmology~\cite{Brandenberger:1988aj} and loop quantum gravity~\cite{Ashtekar:2007em}.

Misner proposed a mixmaster universe where FLRW can emerge at late times from an initial extremely anisotropic, chaotic oscillatory state. Mixmaster phenomenona with bounce and cyclic cosmologies have been widely studied in several Bianchi spacetimes in GR with energy condition-violating fluids \cite{Barrow:1981sx}. In the Bianchi-I class, in particular, bouncing cosmology has been studied in frameworks inspired from string theory \cite{Barrow:1996gx}. A generic feature of Bianchi-I anisotropic bounce in GR is that the effective energy density from spacetime anisotropies grows during the contraction phase as $a^{-6}$ ($a$ is the average scale factor), faster than matter, radiation and spatial curvature, leading to a chaotic mixmaster phenomenon \cite{Misner:1969hg,Barrow:1981sx,Pereira:2007yy}. However, see \cite{Erickson:2003zm,Cai:2013vm,Cook:2020oaj,Ijjas:2020dws} for examples where an Ekpyrotic scenario can control the growth of anisotropies. For Bianchi-IX, it was also noted that quadratic curvature gravity and low energy string theory alleviate the chaotic behavior at initial times \cite{Barrow:1997da}.

However, models of higher derivative gravity are typically plagued by the existence of ghosts \cite{Stelle:1976gc}. One possible way to simultaneously resolve the ghost problem and the cosmological singularity problem is to introduce infinite covariant derivatives in quadratic curvature theories of gravity~\cite{Biswas:2005qr}. Such models of non-local, infinite derivative (field theories and) gravity have been widely studied~\cite{Krasnikov:1987yj}, a class of which has been found to give a stable bounce in FLRW~\cite{Biswas:2005qr,Biswas:2010zk,Biswas:2012bp,Koshelev:2012qn,Koshelev:2013lfm,Kumar:2020xsl}\footnote{In another class of non-local gravity theory super-accelerating bouncing cosmology has been studied at the linearized level \cite{Calcagni:2013vra}. Furthermore, these non-local gravity been studied in the context of cosmic inflation in~\cite{Chialva:2014rla,Craps:2014wga,Koshelev:2016xqb,Koshelev:2017tvv,Koshelev:2020foq,Koshelev:2020xby}, and an emergent FLRW from a string dominated phase in~\cite{Biswas:2006bs}. Quantum aspects of such non-local field theories and gravity were studied in~\cite{Talaganis:2014ida}, and in~\cite{Abel:2019ufz,Abel:2019zou,Abel:2020gdi}, an analogy with the world-line approach to string theory has been explored.}.

Indeed, it is wishful thinking to resolve all the conceivable cosmological singularities at the classical and quantum level, and show that the universe becomes isotropic at late times in order to match with observations. In a recent study, it was shown that the vacuum field equations of non-local gravity containing a GR term, non-local quadratic in Ricci scalar, Ricci tensor and Weyl tensor terms can potentially avoid the BKL type singularity \cite{Koshelev:2018rau}.

In this paper, we restrict ourselves to the study of Bianchi-I type cosmology where there are three scale factors in the three spatial directions, respectively. Our aim is to find a non-singular bouncing solution in a Bianchi-I background in non-local, infinite derivative gravity, whose action includes the GR term (with a cosmological constant) and a non-local quadratic in Ricci scalar term (that is associated with a form factor which is an analytic non-polynomial function of the d'Alembertian $\square$ operator), and a matter sector (which is later determined from the equations of motion). The paper is organized as follows. In Sec.(\ref{sec2}), we present a quick review of the non-local quadratic in Ricci scalar gravity and discuss its cosmological bouncing solutions in isotropic and homogeneous FLRW \cite{Biswas:2005qr,Biswas:2012bp}. In Sec.(\ref{sec3}), we find a new bouncing solution of the theory in an anisotropic but homogeneous Bianchi-I background. In Sec.(\ref{sec4}), we determine the matter sources compatible with the anisotropic bouncing solution found in Sec.(\ref{sec3}) and find conditions on the form factor resulting from requiring positivity of each fluid component's energy density. We show that non-local gravity can ameliorate the challenges faced in a local theory of gravity when it comes to simultaneously satisfying positivity of energy density and absence of instabilities (ghost/tachyonic). In Sec.(\ref{sec5}), we conclude with the main results of our investigation and future outlook. 

Throughout this paper, we fix $\hbar = c = 1$ and follow the metric signature $(-,+,+,+)$.


\section{Lightning review of infinite derivative gravity} \label{sec2}

We will consider a simple form of  non-local generalization of quadratic curvature gravity with a cosmological constant and matter degrees of freedom \cite{Biswas:2005qr}:
\begin{equation} \label{action1}
S = \int d^{4}x\sqrt{-g}\left(   \frac{M_P^{2}}{2} R -  \Lambda  + R \cF (\square) R +\Lc_m  \right)\,,
\end{equation}
where $M_{P}$ is the reduced Planck mass and $\Lc_m$ is the matter Lagrangian. We define dimensionless $\square_{s} \equiv \square / M_s^2 $ and assume the form factor
$\Fc(   \square  )$ to be an analytic function of $\square = g^{\mu \nu} \nabla_{\mu} \nabla_{\nu}$ with a power series expansion\footnote{One may also add non-local terms which are non-analytic, such as $1/\square$~\cite{Deser:2007jk} or $\ln(\square)$ \cite{Barvinsky:2015}. We will not consider them in this paper.}:
\begin{equation} \label{formfactor}
\Fc(   \square  ) =\sum_{n=0}^{\infty}f_{n} \square_{s}^n
\end{equation}
for dimensionless coefficients $f_{n}$. $M_s (<M_P)$ is a new high energy scale of non-locality at which higher derivative terms become important.
The direct bound on $M_s> 0.004$~eV arises from constraints of Newtonian potential at low energies~\cite{Edholm:2016hbt}, while high energy constraints come from gravitational waves and non-Gaussianity~\cite{Koshelev:2020foq}. We will first keep the coefficients in $\cF(\square)$ free. Towards the end of the paper, we will compare local and non-local gravity by choosing specific forms of $\cF(\square)$. Varying the action (\ref{action1}) gives the following equations of motion (EoM) \cite{Biswas:2005qr,Biswas:2010zk,Biswas:2012bp,Biswas:2013cha}
\begin{equation}
\begin{aligned}
& -\left[M_{p}^{2}+4\Fc(\square)R\right]G_{\ \nu}^{\mu}- R\Fc(\square)R\delta_{\ \nu}^{\mu}
+4\left(\nabla^{\mu}\partial_{\nu}-\delta_{\ \nu}^{\mu}\square\right)\Fc(\square)R
 +2\mathcal{K}_{\ \nu}^{\mu}
    -\delta_{\ \nu}^{\mu} (\mathcal{K}_{\ \sigma}^{\sigma} + \tilde{\mathcal{K}})\\
    & - \Lambda \delta^\mu_{\ \nu}
+ T_{\ \nu}^{\mu}=0
\end{aligned}
\label{EoM}
\end{equation}
where $T^\mu_{\ \nu}$ is the energy-momentum tensor and
\begin{equation}
\begin{aligned}\mathcal{K}^{\mu}_{\ \nu}= &  \frac{1}{M_{s}^{2}}\sum_{n=1}^{\infty}f_{n}\sum_{l=0}^{n-1} (\partial^{\mu}\square_{s}^{l}R) (\partial_{\nu}\square_s^{n-l-1}R), \qquad
\tilde{\mathcal{K}} = \sum_{n=1}^{\infty}f_{n}\sum_{l=0}^{n-1} (\square_{s}^{l}R) (\square_{s}^{n-l}R).
\end{aligned}
\label{KmnRFR-1}
\end{equation}
The trace EoM is
\begin{equation}
(M_P^2 -12\square\Fc(  \square )) R- 2\Kc^\mu_{\ \mu} - 4 \tilde{\Kc} -4\Lambda= -T.
\label{trace}
\end{equation}
Evidently, solving the EoM (\ref{EoM}) in general is very non-trivial. However, they can be greatly simplified if the background follows some simplifying (but motivated) assumptions. The complexity of non-local EoM due to an infinite tower of $\square$ operators acting on curvatures can be reduced drastically if we take the background Ricci scalar to satisfy the following ansatz\footnote{The EoM of non-local, infinite derivative gravity have been exactly solved recently with a different ansatz within the class of almost universal spacetimes \cite{Dengiz:2020xbu,Kolar:2021rfl}.} \cite{Biswas:2005qr,Biswas:2010zk,Biswas:2012bp}
\be \label{ansatz}
\square R= r_1R+r_2
\ee
where $r_{1,2}$ are dimensionful constants. Using this ansatz, we can now reduce the number of $\square$ operators acting on curvatures in the EoM (\ref{EoM}), since
\be
\square^n R= r_1^n \left(\bar{R}+\frac{r_2}{r_1} \right) \qquad \text{and} \qquad \Fc (\square) R = \Fc_1 \bar{R}+\Fc_2
\ee
where we have defined
\begin{equation} \label{fdefs}
\cF_{1} \equiv \cF (r_{1}) \qquad \text{and}  \qquad  \Fc_2 \equiv \frac{r_2}{r_1} ( \Fc_1 - f_{0} ).
\end{equation}
Substituting the ansatz (\ref{ansatz}) in the EoM (\ref{EoM}), we obtain a simple expression for the energy-momentum tensor \cite{Biswas:2010zk,Biswas:2012bp}
\be \label{fullT}
\ba
T^{\mu}_{\ \nu} =
& \left[M_P^2+4 \left(\Fc_1 R + \cF_{2} \right)\right] G^\mu_{\ \nu}
+ \Lambda \delta^\mu_{\ \nu}
- 2 \Fc'_{1} \left[\pd^\mu R \pd_\nu R-\frac{\delta^\mu_{\ \nu}}{2}\left(g^{\sigma\rho}\pd_\sigma R \pd_\rho R+r_1\left(R+\frac{r_2}{r_1}\right)^2\right)\right]\\
&\quad  -4 \Fc_1\left[\nabla^\mu\pd_\nu R-\delta^\mu_{\ \nu}(r_1R+r_2)\right]
+\delta^\mu_{\ \nu} \left[\Fc_1R^2-\frac{r_2}{r_1} \Fc_2 \right]
\ea
\ee
where $\Fc'_{1} \equiv \Fc' (r_1)$ denotes the first derivative of $\cF(\square)$ with respect to $\square$, evaluated at $\square = r_{1}$. In particular, for traceless matter with $T^\mu_{\ \mu}=0$, substituting the ansatz (\ref{ansatz}) in the trace EoM (\ref{trace}) gives us the following unique conditions on $\cF(\square)$ \cite{Biswas:2010zk,Biswas:2012bp}
\begin{equation}
\Fc'_{1} = 0,\qquad \Fc_2= -\frac{M_p^2}{4}+3 r_{1} \Fc_1 ,\qquad  \Lambda = -\frac{M_P^2}{4} \frac{r_2}{r_1}
\label{condi}
\end{equation}
which substituting in (\ref{fullT}) provides an even simpler expression for the energy-momentum tensor
\begin{equation}
T^\mu_{\ \nu} = 4\Fc_1 \Big[(  R+3r_1 ) G^\mu_{\ \nu} -\nabla^\mu\partial_\nu R +\frac{1}{4} \delta^\mu_{\ \nu} (  R^2+4r_1 R+r_2 ) \Big].
\label{tmnan}
\end{equation}
Everything discussed until now is covariant, that is, we did not choose any specific metric. In this paper, we will consider cosmological spacetimes, particularly bouncing scenarios. As a specific example, for spatially flat, homogeneous and isotropic FLRW, the scale factor~\cite{Biswas:2005qr} 
\begin{equation}\label{coshbounce}
a(t) = a_{0} \cosh \left(  \sqrt{\frac{r_{1}}{2}} t \right)\,,
\end{equation}
where $a_0$ is the scale factor at the bounce point $t=0$, describes a bouncing universe which solves the ansatz (\ref{ansatz}) with $r_{2} = -6 r_{1}^{2}$ \cite{Biswas:2005qr,Biswas:2010zk,Biswas:2012bp,Koshelev:2012qn}. It is therefore also a solution of the EoM (\ref{EoM}) of quadratic curvature non-local gravity (\ref{action1}). In the case of traceless matter, this $\cosh$ bounce is an exact solution of the EoM (\ref{tmnan}) with an energy density arising from radiation and given by $\rho_{r} = -27 \cF_{1} r_{1}^{2}$, which implies $\cF_{1}<0$ if we want positive energy density $\rho_{r}>0$ for stability \cite{Biswas:2012bp}.

In a local $R+f_{0} R^{2} - \Lambda$ theory, $\cF_{1} = f_{0}$ and $f_{0}<0$ leads to a tachyonic instability  \cite{Schmidt:2006jt}. In non-local gravity, on the other hand, it is possible to fix $\cF_{1}<0$ without leading to any instabilities. Note that the cosh solution approaches de Sitter (dS) asymptotically at early ($t \to -\infty$) and late ($t \to +\infty$) times, with a constant Ricci scalar $R = 6 r_{1}$\footnote{Another interesting bouncing solution (in spatially flat FLRW) of the ansatz (\ref{ansatz}), and hence of quadratic curvature non-local gravity (\ref{action1}), is $a (t) = a_{0} e^{\frac{\lambda}{2}t^{2}}$ with $r_{1} = -6 \lambda$, $r_{2} = 12 \lambda^{2}$ and $\Lambda =\lambda M_p^2/2$ \cite{Koshelev:2012qn}. It is worth noting here that bouncing cosmology has also been studied in the context of Deser-Woodard infrared non-local modification of gravity \cite{Chen:2019wlu}.}.
%


\section{Bouncing solutions in anisotropic backgrounds}

\label{sec3}

As just mentioned, cosmological bouncing solutions which satisfy the ansatz (\ref{ansatz}) are already known when the background is flat, homogeneous and isotropic (FLRW). But a bounce need not happen in an isotropic manner. Therefore, we choose a bit more complicated spacetime such as the homogeneous but anisotropic Bianchi-I background. The Bianchi-I spacetime metric can in general be expressed as \cite{Pereira:2007yy}
\begin{equation}
ds^2=-dt^2+a^2 \left[ e^{2\beta_1}dx_1^2 + e^{2\beta_2}dx_2^2 + e^{2\beta_3}dx_3^2 \right]
\label{B-Im}
\end{equation}
where $a = a(t)$ is the average scale factor (geometric mean of scale factors in the three spatial directions), and where deviation from FLRW is parameterized by $\beta_{i} = \beta_{i} (t)$ (in each spatial direction $i=1,2,3$) which satisfy
\be \label{betacond}
\beta_1+\beta_2+\beta_3=0.
\ee
%
%
%
%
A quantity called shear $\sigma^2$ defined as
\begin{equation}
\sigma^2=\dot{\beta}_1^2+\dot{\beta}_2^2+\dot{\beta}_3^2\,,
\label{shearv}
\end{equation}
essentially parameterizes the extent of anisotropic deviation of the background away from FLRW.
The Ricci scalar for the anisotropic metric (\ref{B-Im}) is
\begin{equation}
R=12H^2+6\dot{H}+\sigma^2 = \bar{R} + \sigma^{2}\,,
\label{Ricci-newv}
\end{equation}
where $\bar{R}$ is the Ricci scalar for an FLRW metric with the scale factor $a$, and $H= \dot{a}/a$ is the average of the three Hubble factors in the three spatial directions. Particularly, from the metric (\ref{B-Im}), we can define the three scale and Hubble factors in the three spatial directions ($i=1,2,3$) as 
\begin{equation} \label{spatialhubblefactors}
a_{i} = a e^{\beta_{i}},~~ H_{i} = H + \dot{\beta}_{i}\,,~~~H=\frac{1}{3}(H_{1}+H_{2}+H_{3})\,.
\end{equation}
The $\square$ operator for the Bianchi-I metric (\ref{B-Im}) acting on a time-dependent scalar is
\begin{equation}
\square =  -\frac{\pd^2}{\pd t^2}-3H\frac{\pd}{\pd t}\,,
\label{Box-newv}
\end{equation}
which is exactly the same as in FLRW. In this paper, we generalize the known $\cosh$ bounce in FLRW to Bianchi-I backgrounds (\ref{B-Im}). We take the average scale factor to be $\cosh$, see (\ref{coshbounce}), where $a_{0}$ is now the average scale factor at the bounce point $t=0$. Henceforth, we choose a convenient normalization of $a_{0} = 1$ in  (\ref{coshbounce}) so that the minimum value of the average scale factor is $1$. Just as in FLRW, we assume the following ansatz to hold for the Bianchi-I metric (\ref{B-Im})
\be \label{ansatz2}
\square R = s_{1} R + s_{2}\,,
\ee
where $s_{1,2}$ are some dimensionful constants. Substituting (\ref{Ricci-newv}) in (\ref{ansatz2}), we obtain $s_{1} = r_{1}, s_{2} = r_{2}$ (where $r_{1,2}$ are the dimensionful constants in FLRW $\cosh$ bounce with $\square \bar{R}= r_1\bar{R}+r_2$) and the following second order differential equation
\be
\square \sigma^{2} = s_{1} \sigma^{2}\,,
\ee
which can be solved to obtain the shear as
\be \label{shearcosh}
\sigma^{2} =  \sigma_{0}^{2} \ \text{sech}^{2}\left(  \sqrt{\frac{r_{1}}{2}} t \right) = \frac{\sigma_{0}^{2}}{a^{2}}
\ee
subject to boundary conditions $\sigma^{2}|_{t=0} = \sigma_{0}^{2}$ and $\frac{d}{dt}(\sigma^{2})|_{t=0} = 0$. The first condition is normalization, while the second is required to obtain positive $\sigma^{2}$.

In GR, shear goes like $\sigma^{2} \sim a^{-6}$, and the effective energy density coming from the anisotropic background $\rho_{\sigma} = \sigma^{2}/2$ therefore dominates over matter, radiation and spatial curvature when the scale factor $a$ becomes very small \cite{Battefeld:2014uga}. This leads the geometry during the contraction phase into a Kasner solution, and also causes BKL instability \cite{Belinsky:1970ew}.

On the other hand, in a cosh bounce (\ref{coshbounce}) satisfying the ansatz (\ref{ansatz2}) in a Bianchi-I universe (\ref{B-Im}), $\sigma^{2} \sim a^{-2}$ increases much slower in comparison during contraction, reaches a maximum at the bounce point $t=0$, and then decreases during the expansion phase (and, moreover, exponentially as $\sim e^{-2t}$ at late times).
From (\ref{shearv}) and (\ref{shearcosh}), we obtain (for $i=1,2,3$)
\be \label{betasol}
\dot{\beta}_{i} = \sigma_{0} b_{i} \ \text{sech}  \left( \sqrt{\frac{r_{1}}{2}} t \right) \qquad \text{or} \qquad \beta_{i} = \sqrt{\frac{2}{r_{1}}} \sigma_{0} b_{i} \ \text{gd} \left( \sqrt{\frac{r_{1}}{2}} t \right)
\ee
where $b_{i}$ are dimensionless coefficients and $\text{gd} (t)$ is the Gudermannian function which is defined as gd(t) $= \int_{0}^{t} dx \ \text{sech($x$)}$. Any constants of integration can be absorbed into coordinate redefinitions. 
%
%
%
\begin{figure}
	\centering
	\includegraphics[width=0.55\linewidth]{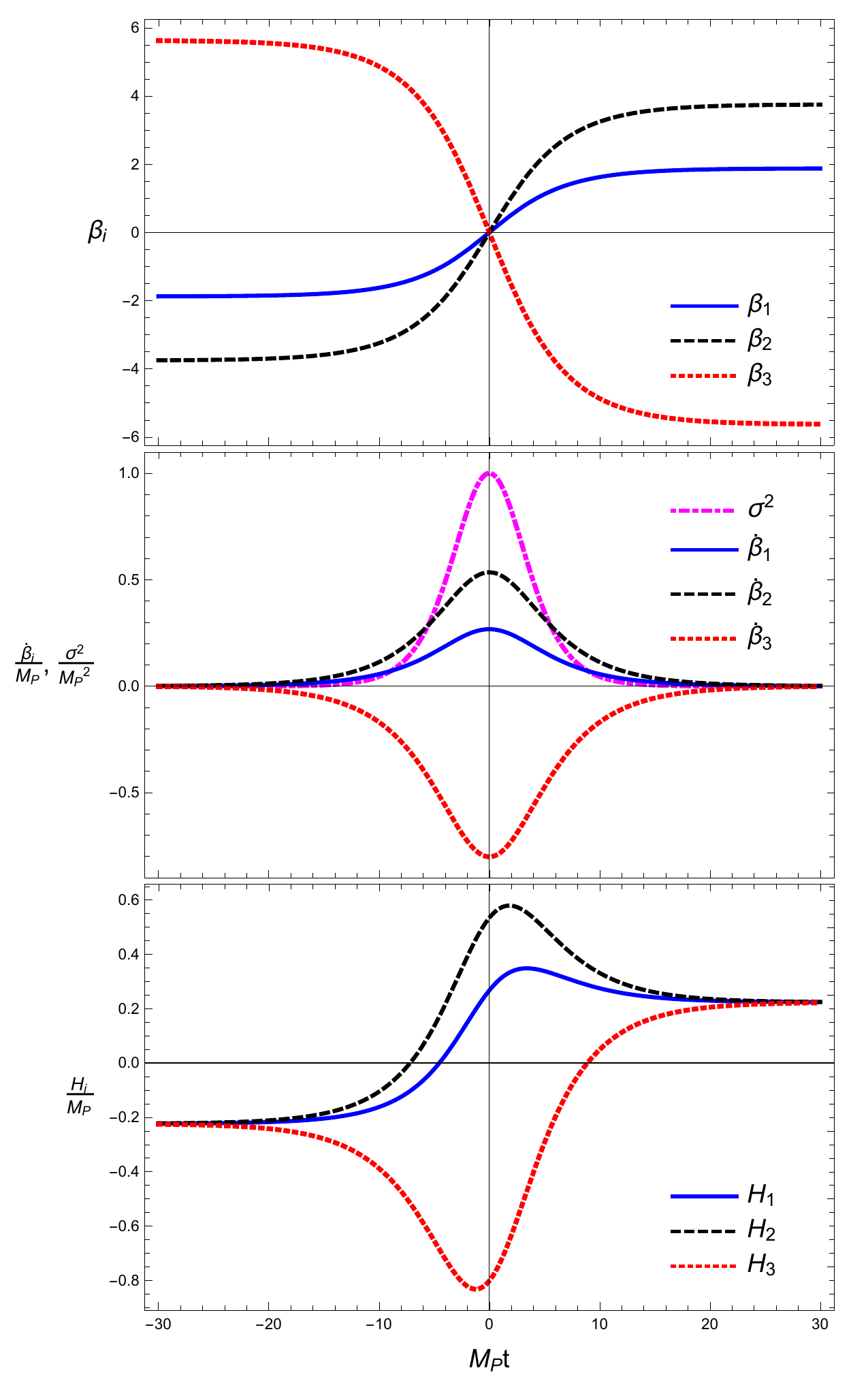}
	\caption{Time evolution (in corresponding units of $M_{P}$) of anisotropy factors $\beta_{i}(t)$ (top plot) and $\dot{\beta}_{i}(t)/M_{P}$ (middle plot) given in (\ref{betasol}), shear $\sigma^{2}(t)/M_{P}^{2}$ given in (\ref{shearcosh}) (middle plot), and Hubble factors in the three spatial directions $H_{i}/M_{P}$ given in (\ref{spatialhubblefactors}) (bottom plot). We have chosen $r_{1}=M_s^2=0.1 M_{P}^{2}$ and $\sigma_{0}=1 M_{P}$ such that the universe is dominated by anisotropy as $t\to 0$. We chose $(b_{1},b_{2},b_{3}) = \left(\frac{1}{\sqrt{14}}, \frac{2}{\sqrt{14}}, \frac{-3}{\sqrt{14}}\right)$ in accordance with (\ref{betaconds}). In the middle plot, we see that the magnitude of background anisotropy (in each spatial direction $i=1,2,3$) reaches a maximum at the bounce point $t=0$, and then asymptotes to zero before and after, as the universe asymptotes to FLRW (and eventually to dS, as can be seen from the bottom plot where $H_{i}$ approach constant values for $t \gg t_{\rm iso}$ (see (\ref{tisotime}))). The cosmological constant $\Lambda = \frac{3}{2} M_{P}^{2} M_{s}^{2}$ (see (\ref{lambdaval})). In models based on GR, the shear $\sigma^2$ decays like $a^{-6}$ leading to strong growth of anisotropies during the contraction phase \cite{Battefeld:2014uga}. On the other hand, in non-local quadratic scalar curvature theory (\ref{action1}), since $\sigma^2$ evolves like $a^{-2}$, we can have a smoother transition from contraction to expansion.}
	\label{fig:betaandbetadot}
\end{figure}
%
From (\ref{betacond}) and (\ref{shearv}), we get the following constraints
\be \label{betaconds}
b_{1} + b_{2} + b_{3} = 0, \qquad b_{1}^{2} + b_{2}^{2} + b_{3}^{2} = 1
\ee
which admit an infinite number of solutions for the triplet $(b_{1},b_{2},b_{3})$.

 Background anisotropy is quantified by the shear $\sigma^2(t)$  which from (\ref{shearcosh}) becomes maximum at the bounce point $t=0$, and goes to zero exponentially as $t \to \pm \infty$ (see Fig.(\ref{fig:betaandbetadot})). The universe therefore approaches isotropic and homogeneous FLRW asymptotically as $t \to \pm \infty$, along with entering a dS phase. We can understand better the isotropic and anisotropic limits of spacetime by looking at the structure of the Ricci scalar given in \eqref{Ricci-newv}, where we can compare the FLRW $\bar{R}$ part of the full Ricci scalar $R$ that depends on the average Hubble parameter $H$ with the part corresponding to the anisotropic shear $\sigma^2$.  We can say that the spacetime anisotropy-dominated phase of evolution is when $\sigma^2(t) \gg \bar{R}$ (i.e., when the maximum contribution to the curvature scalar $R$ comes from the shear) while it is isotropy-dominated when $\sigma^2(t) \ll \bar{R}$ (i.e., when the average evolution dominates over anisotropies).  In Fig.~(\ref{fig:betaandbetadot}), we depict the evolution of the universe which is extremely anisotropic towards $t\to 0$, reaching towards isotropy in the limit $t\to \pm \infty$. We can determine the time scale $t_{\rm iso}$ at which the universe starts turning towards isotropic evolution as follows 
 \begin{equation} \label{tisotime}
 \bar{R} \sim \sigma^2 \implies	t_{\rm iso} \sim \pm
 	\sqrt{\frac{2}{r_1}}\cosh^{-1}\left(\sqrt \frac{3r_1+\sigma_0^2}{6r_1}\right)\,. 
 \end{equation}
Note that for the quadratic curvature theory in (\ref{action1}) with the non-locality scale $M_{s}$, we can choose\footnote{This choice is just for simplicity. An exact expression for $r_{1}$ in terms of $M_{s}^{2}$ cannot be found just from the EoM of the non-local theory in (\ref{fullT}) because from there, due to the presence of the form factor $\cF(\square)$ defined in (\ref{formfactor}), one obtains (in general) an infinite order algebraic equation for $r_{1}$. Moreover, for an arbitrary $\cF(\square)$, there are an infinite number of undetermined coefficients $f_{n}$ in its power series expansion, preventing us from uniquely fixing $r_{1}$ in terms of $M_{s}^{2}$.} $r_{1} = M_{s}^{2}$ as in Fig.(\ref{fig:betaandbetadot}). For the values chosen in Fig.(\ref{fig:betaandbetadot}), $r_{1}=0.1 M_{P}^{2}$ and $\sigma_{0}=1 M_{P}$, we obtain $t_{iso} \approx \pm 4.19 M_{P}^{-1}$ at which point the shear drops to $\sigma^{2}(t = t_{iso}) \approx 0.46 M_{P}^{2}$, down from its maximum value of $1 M_{P}^{2}$ at $t=0$. As time goes on, the shear drops to $\sigma^{2}(t = 2 t_{iso}) \approx 0.09 M_{P}^{2}$ and $\sigma^{2}(t = 4 t_{iso}) \approx 0.002 M_{P}^{2}$.


\section{Matter sources for anisotropic bounces}
\label{sec4}

Having completely fixed the anisotropic bouncing background in the previous section, we now determine the corresponding possible matter sectors which can drive such a non-singular evolution of the universe (\ref{coshbounce}) with controlled growth of anisotropies (\ref{betasol}). We show how anisotropic shear $\sigma^{2}$ effectively contributes to fluid energy densities and pressures. We also determine the conditions where all component energy densities are non-negative, ensuring a physically healthy bounce mechanism. Moreover, we also show that under certain conditions, it is possible to have a bounce in vacuum (but with $\Lambda$).

Without loss of generality, the energy-momentum tensor $T^{\mu}_{\ \nu}$ for quadratic curvature non-local gravity (\ref{action1}) around a Bianchi-I cosmological background (\ref{B-Im}) has the diagonal form
\be
T^{\mu}_{\ \nu} = (-\rho, p_{1}, p_{2}, p_{3})
\ee
where $p_{i}$ (for $i=1,2,3$) are pressures in $x,y,z$ directions. We will soon see that anisotropic pressures necessarily arise in some (but not all) cases of possible matter sectors. For a perfect fluid, $p_{1} = p_{2} = p_{3}$ and the energy density of each component evolves as
\be
\rho_{I} = \rho_{I0} \ {a}^{-3(1+w_{I})}
\ee
where $I$ labels the different species, $\rho_{I0}$ denotes the energy density at the bounce point of component $I$, and $w_{I}$ is its equation of state. The zero component of energy-momentum conservation $\nabla_{\mu} T^{\mu}_{\ \nu}=0$ in a Bianchi-I background (\ref{B-Im}) takes the form
\be \label{emcons}
\dot{\rho} + 3 H \rho + H(p_{1}+p_{2}+p_{3}) + \dot{\beta}_{1} p_{1}+ \dot{\beta}_{2} p_{2}+ \dot{\beta}_{3} p_{3} = 0.
\ee
We now find the energy-momentum tensor responsible for the bounce. We use the EoM in (\ref{fullT}) and substitute in it the cosh bouncing solution in a Bianchi-I background (\ref{B-Im}), (\ref{coshbounce}), (\ref{shearcosh}) and (\ref{betasol}) to obtain an energy density with three components
\be
\ba \label{rhofullcomponents}
\rho
= \frac{\rho_{k0}}{a^{2}} + \frac{\rho_{r0}}{a^{4}} + \frac{\rho_{s0}}{a^{6}}
\ea
\ee
where
\be \label{rhofull}
\ba
\rho_{k0} &= - \frac{1}{2} \sigma_{0}^2 \left(60 \cF_{1}   r_{1} +4 \cF_{2} +M_P^2\right) -   \frac{3 r_{1}}{2}  \left(-12 \cF_{1}   r_{1} +4 \cF_{2} +M_P^2\right) \\
\rho_{r0} &= -3 \left(\sigma_{0}^2-3  r_{1} \right)^2 (\cF_{1} +\cF'_{1}  r_{1} ) \\
\rho_{s0} &= 2 \cF'_{1}  r_{1}  \left(\sigma_{0}^2-3  r_{1} \right)^2
\ea
\ee
where the first component is `$k$-matter'  whose energy density scales as $a^{-2}$ (just like spatial curvature in GR), second is radiation labeled by `$r$', and the last is stiff matter labeled by `$s$' whose energy density scales as $a^{-6}$. We see that anisotropic shear, parameterized by $\sigma_{0}$, effectively contributes to all the three matter sectors. Note that the cosmological constant $\Lambda$ is fixed to be
\be \label{lambdaval}
\Lambda = \frac{3}{2} M_P^2  r_{1}
\ee
in order to absorb the constant term in $\rho$ in (\ref{rhofullcomponents}). Pressure is anisotropic in general (due to $k$-matter) and given by (in each spatial direction $i=1,2,3$)
\be
\ba
p_{i} = \left( \frac{p_{k0}}{a^{2}} + \pi^{i}_{\ i} \right) + \frac{p_{r0}}{a^{4}} + \frac{p_{s0}}{a^{6}}
\ea
\ee
where
\be \label{pressures}
\ba
p_{r0} &= w_{r} \rho_{r0} \quad \text{with} \quad w_{r} = 1/3, \qquad \qquad
p_{s0} = w_{s} \rho_{s0} \quad \text{with} \quad w_{s} = 1,\\
p_{k0} &= w_{k} \rho_{k0} \ \text{with} \ w_{k} = \frac{- \frac{1}{2} \sigma_{0}^2 \left(12 \cF_{1}   r_{1} +4 \cF_{2} +M_P^2\right) +   \frac{r_{1}}{2}  \left(-12 \cF_{1}   r_{1} +4 \cF_{2} +M_P^2\right)}{- \frac{1}{2} \sigma_{0}^2 \left(60 \cF_{1}   r_{1} +4 \cF_{2} +M_P^2\right) -   \frac{3 r_{1}}{2}  \left(-12 \cF_{1}   r_{1} +4 \cF_{2} +M_P^2\right)},
\ea
\ee
and the $k$-matter anisotropic pressure $\pi^{i}_{\ i}$ (in each spatial direction $i=1,2,3$) is given by
\be \label{anisopressure}
\pi^{i}_{\ i} = \pm \sqrt{2 r_{1}} b_{i} \sigma_{0} (4 \cF_{2} + M_{P}^{2} + 24 \cF_{1} r_{1}) \sqrt{\frac{1}{a^{2}} - \frac{1}{a^{4}}} \qquad (+ \text{ for } t>0, \ - \text{ for } t<0)
\ee
where there is no summation over the index $i$ in $\pi^{i}_{\ i}$. The appearance of $\pi^{i}_{\ i}$ is a consequence of purely $\dot{\beta}_{i}$ and $\ddot{\beta}_{i}$-dependent terms in the EoM\footnote{In GR, one obtains a second order differential equation for $\beta_{i}$ by using the Friedmann equations and the constraint on $\beta_{i}$ (\ref{betacond}) in the presence of a perfect fluid matter sector. Essentially, the purely $\dot{\beta}_{i}$ and $\ddot{\beta}_{i}$-dependent terms in $T^{i}_{\ i} (=G^{i}_{\ i})$ put to zero gives the sought-after differential equation. In our case, we already determined $\beta_{i}$ (\ref{betasol}) by solving the Ricci scalar ansatz (\ref{ansatz2}) for a fixed average scale factor (\ref{coshbounce}) in a Bianchi-I background. However, this $\beta_{i}$ does not put the purely $\dot{\beta}_{i}$ and $\ddot{\beta}_{i}$-dependent terms in $T^{i}_{\ i}$ (\ref{fullT}) to zero, as is evident from non-zero $\pi^{i}_{\ i}$ (\ref{anisopressure}).}. The $\sigma^{2}$-dependent contributions to the energy density and pressure, on the other hand, are isotropic in all spatial directions. This shows that, in general, the matter sector consists of two perfect fluids (radiation and stiff matter) and one fluid with anisotropic pressures ($k$-matter). Note from (\ref{pressures}) that $k$-matter, in general, does not have the correct, constant equation of state $w_{k}=-1/3$ for it to be a perfect fluid, necessitating anisotropic pressures $\pi^{i}_{\ i}$ in order to ensure energy-momentum conservation (\ref{emcons}). In particular, we have the energy-momentum tensors of perfect fluid and $k$-matter as, respectively,
\be
\ba
T^{I}_{\mu \nu} = (p_{I}+\rho_{I}) u_{\mu} u_{\nu} + p_{I} g_{\mu \nu}, \qquad
T^{k}_{\mu \nu} = (p_{k}+\rho_{k}) u_{\mu} u_{\nu} + p_{k} g_{\mu \nu} + \pi_{\mu \nu}
\ea
\ee
where $I=r$ or $s$, and $u_{\mu}$ is the velocity vector.
The anisotropic pressure contribution $\pi^{\mu}_{\ \nu} = (0,\pi^{1}_{\ 1},\pi^{2}_{\ 2},\pi^{3}_{\ 3})$ is traceless $\pi^{\mu}_{\ \mu}=0$ and transverse to the velocity vector $\pi_{\mu \nu} u^{\mu} = 0$. Background anisotropy, parameterized by $\sigma_{0}$, does not introduce any new kinds of (effective) matter apart from those already present in the known FLRW cosh bouncing model. The FLRW limit can be taken by putting $\sigma_{0} \to 0$ in (\ref{rhofull}), (\ref{pressures}) and (\ref{anisopressure}) \cite{Biswas:2005qr}. We enumerate all possible matter sectors in Table (\ref{table1}), where the constraints on $\cF_{1}, \cF'_{1},\cF_{2}$ and $\sigma_{0}$ are determined by requiring that the energy density of each matter component be non-negative. In particular, requiring $\rho_{s0} \geq 0, \rho_{r0} \geq 0$ and $\rho_{k0} \geq 0$ constrains $\cF'_{1},\cF_{1}$ and $\cF_{2}$, respectively (see (\ref{rhofull})).
In some special cases, however, upon constraining the parameters of the theory and solution, it is possible to turn $k$-matter into a perfect fluid. From (\ref{anisopressure}), we see that imposing a constraint
\be \label{pfluidcond}
4 \cF_{2} + M_{P}^{2} + 24 \cF_{1} r_{1} = 0
\ee
makes anisotropic pressures vanish $\pi^{i}_{\ i}=0$, and recovers the correct, constant equation of state for $k$-matter $w_{k}=-1/3$, leaving us with three perfect fluids in the matter sector. Note that (for $\cF_{1} \neq 0$) it is impossible to satisfy both the perfect fluid constraint (\ref{pfluidcond}) and the second condition in (\ref{condi}) for traceless matter. This is another way of saying that a traceless matter sector driving an anisotropic $\cosh$ bounce must necessarily have anisotropic pressures (in the $k$-matter component). In Appendix (\ref{appendix}), we determine the potential for $k$-matter, assuming that it can be described by a perfect fluid scalar field.



\subsection{Most general anisotropic bounce}

Consider as an example the most general case when all three matter components $(k,r,s)$ are present in the model. A priori, $\cF_{1}, \cF'_{1},\cF_{2},r_{1}$ and $\sigma_{0}$ are independent of each other. Relations between them are derived by invoking physical principles like positivity of energy density. Imposing this positivity for each component in (\ref{rhofull}), $\rho_{s0} > 0$, $\rho_{r0} > 0$ and $\rho_{k0} > 0$, we obtain the following conditions to be satisfied by $\cF'_{1},\cF_{1}$ and $\cF_{2}$, respectively (see also Table (\ref{table1}))
\be \label{condi2}
\ba
\Fc'_{1} > 0, \qquad \cF_{1} < - r_{1} \Fc'_{1} < 0, \qquad
 \Fc_{2} < -\frac{M_{P}^2}{4} + 3 \cF_{1} r_{1} \left(1 - \frac{6 \sigma _0^2}{3 r_{1} + \sigma _0^2}\right).
\ea
\ee
Upon imposing the perfect fluid condition on $\cF_{2}$ in (\ref{pfluidcond}), the energy density of $k$-matter at the bounce point becomes
\be
\rho_{k0} = 18 \cF_{1} r_{1} \left(3 r_{1} - \sigma_{0}^2 \right)
\ee
which in turn implies a lower bound on background anisotropy $\sigma_{0}^{2} > 3 r_{1}$ if we wish to have $\rho_{k0} > 0$. No such bound on $\sigma_{0}$ exists in the general case with two perfect radiation and stiff matter fluids and one anisotropic $k$-matter fluid.

Let us go back to the most general case with three fluids (\ref{rhofull}) and (\ref{condi2}). Interestingly, for the specific value $\sigma_{0}^{2} = 3 r_{1}$, radiation and stiff matter vanish, leaving only anisotropic $k$-matter with the following energy density at the bounce point
\be
\rho_{k0} = - \sigma_{0}^{2} (4 \cF_{2} + M_{P}^{2} + 8 \cF_{1} \sigma_{0}^{2})
\ee
which implies
\be
 \Fc_{2} < -\frac{M_{P}^2}{4} - 2 \cF_{1} \sigma_{0}^{2}
\ee
if we require $\rho_{k0}>0$. Moreover, if we now impose the perfect fluid condition (\ref{pfluidcond}) on $\cF_{2}$ with this fixed value $\sigma_{0}^{2} = 3 r_{1}$, the energy densities of all fluids vanish and we obtain an anisotropic bounce in vacuum (but with $\Lambda$) (see Table (\ref{table1})).

There exists another set of conditions where the bounce is in vacuum, but now with unconstrained $\sigma_{0}^{2}$. Consider again the most general case (\ref{rhofull}). We fix $\cF'_{1} = \cF_{1} = 0$ to remove stiff matter and radiation from the model, and obtain
\be
\rho_{k0} = -\frac{1}{2} (4 \cF_{2}+M_{P}^2) (3 r_{1}+\sigma_{0}^2)
\ee
which implies that $\cF_{2} = -\frac{M_{P}^{2}}{4}$ if we want to have a bounce in vacuum. This vacuum solution is the same one we would obtain if we start from a traceless matter theory, already satisfying the tracelessness conditions in (\ref{condi}), and then put $\cF_{1}=0$ in (\ref{tmnan}).
These two vacuum solutions identify with each other for one specific tuning $\cF'_{1} = \cF_{1} = 0$, $\cF_{2} = -\frac{M_{P}^{2}}{4}$ and $\sigma_{0}^{2} = 3 r_{1}$. They are mutually exclusive otherwise.


\begin{table}[h!]
\begin{center}
  \begin{tabular}{c | c | c | c | c | c | c }
    \hline \hline
     $a^{-6}$ & $a^{-4}$ & $a^{-2}$ & $\cF_{1}$ & $\cF'_{1}$ & $\cF_{2}$   &Perfect fluid possible?\\ \hline
     \color{green}\cmark & \color{green}\cmark & \color{green}\cmark   &$<-r_{1} \cF'_{1}<0$ &$>0$ 	&$< -\frac{M_{P}^2}{4} + 3 \cF_{1} r_{1} \left(1 - \frac{6 \sigma _0^2}{3 r_{1} + \sigma _0^2}\right)$        & if (\ref{pfluidcond}) is true; $\sigma_{0}^{2} > 3 r_{1}$     \\ \hline
          \color{red}\xmark & \color{green}\cmark & \color{green}\cmark   &$<0$ &$0$ 	&$< -\frac{M_{P}^2}{4} + 3 \cF_{1} r_{1} \left(1 - \frac{6 \sigma _0^2}{3 r_{1} + \sigma _0^2}\right)$        & if (\ref{pfluidcond}) is true; $\sigma_{0}^{2} > 3 r_{1}$     \\ \hline
          \color{green}\cmark & \color{red}\xmark & \color{green}\cmark   &$- r_{1} \cF'_{1}<0$ &$>0$ 	&$< -\frac{M_{P}^2}{4} + 3 \cF_{1} r_{1} \left(1 - \frac{6 \sigma _0^2}{3 r_{1} + \sigma _0^2}\right)$        & if (\ref{pfluidcond}) is true; $\sigma_{0}^{2} > 3 r_{1}$     \\ \hline
               \color{green}\cmark & \color{green}\cmark & \color{red}\xmark   &$<-r_{1} \cF'_{1}<0$ &$>0$ 	&$ -\frac{M_{P}^2}{4} + 3 \cF_{1} r_{1} \left(1 - \frac{6 \sigma _0^2}{3 r_{1} + \sigma _0^2}\right)$        & always     \\ \hline
                    \color{red}\xmark & \color{red}\xmark & \color{green}\cmark   &$0$ &$0$ 	&$< -\frac{M_{P}^2}{4} $        & never     \\ \hline
                                        \color{red}\xmark & \color{green}\cmark & \color{red}\xmark   &$<0$ &$0$ 	&$ -\frac{M_{P}^2}{4} + 3 \cF_{1} r_{1} \left(1 - \frac{6 \sigma _0^2}{3 r_{1} + \sigma _0^2}\right)$        & always     \\ \hline
        \color{green}\cmark & \color{red}\xmark & \color{red}\xmark   &$-r_{1} \cF'_{1}<0$ &$>0$ 	&$ -\frac{M_{P}^2}{4} + 3 \cF_{1} r_{1} \left(1 - \frac{6 \sigma _0^2}{3 r_{1} + \sigma _0^2}\right)$        & always     \\ \hline
        \color{red}\xmark & \color{red}\xmark & \color{green}\cmark   &- &-	& $< -\frac{M_{P}^2}{4} - 2 \cF_{1} \sigma_{0}^{2}$        & never;  $\sigma_{0}^{2}=3r_{1}$     \\ \hline
         \color{red}\xmark & \color{red}\xmark & \color{red}\xmark   &- &-	& see (\ref{pfluidcond})        & vacuum;  $\sigma_{0}^{2}=3r_{1}$     \\ \hline
         \color{red}\xmark & \color{red}\xmark & \color{red}\xmark   &$0$ &$0$	& $-\frac{M_{P}^2}{4}$        & vacuum     \\ \hline
    \hline
  \end{tabular}
\end{center}
\caption{Possible matter sectors for a bounce in a Bianchi-I background for the quadratic curvature theory (\ref{action1}) following the EoM in (\ref{fullT}) and the background Ricci scalar ansatz (\ref{ansatz2}). At the top of the first three columns, $a^{-6}$ represents stiff matter, $a^{-4}$ represents radiation and $a^{-2}$ represents $k$-matter. Conditions on $\cF_{1},\cF'_{1}$ and $\cF_{2}$ arise from imposing non-negativity of energy density of each matter component, $\rho_{r0} \geq 0, \rho_{s0} \geq 0$ and $\rho_{k0} \geq 0$, respectively (see (\ref{rhofull})). Note that the inequality for $\cF_{2}$ can be rewritten as an inequality for $f_{0}$ instead, using the definition of $\cF_{2}$ in (\ref{fdefs}). A hyphen `-' indicates that the value is unconstrained. The value of background anisotropy $\sigma_{0}$ is unconstrained, unless mentioned otherwise. The last column enumerates the cases when $k$-matter (and therefore the entire matter sector) can be a perfect fluid.}
\label{table1}
\end{table}


It is instructive to mention the cosh bouncing model (\ref{coshbounce}) in a Bianchi-I background (\ref{B-Im}) when the matter sector is traceless. For this purpose, we substitute the tracelessness conditions (\ref{condi}) in the most general EoM (\ref{fullT}) to obtain the simplified EoM (\ref{tmnan}). In FLRW, the only matter component is radiation \cite{Biswas:2010zk}. In a Bianchi-I background, however, we see that background anisotropy $\sigma_{0}$ introduces $k$-matter, effectively. Upon imposing the tracelessness conditions (\ref{condi}) in the general expression for $\rho_{k0}$ and $p_{k0}$ in (\ref{rhofull}) and (\ref{pressures}), we obtain $\rho_{k0} = - 36 \cF_{1} r_{1}  \sigma_{0}^{2}$ and $w_{k}=1/3$. Note that the condition $\cF'_{1}=0$ in (\ref{condi}) removes stiff fluid from the model. The equation of state corresponds to radiation, even though the fluid has an energy density which decays like $a^{-2}$. This necessitates anisotropic pressures $\pi^{i}_{\ i}$ in order to conserve energy-momentum (\ref{emcons}), which are given by
\be
\pi^{i}_{\ i} = \pm 36 b_{i} \cF_{1} \sigma_{0}  \sqrt{2 r_{1}^{3}} \sqrt{\frac{1}{a^{2}} - \frac{1}{a^{4}}} \qquad (+ \text{ for } t>0, \ - \text{ for } t<0).
\ee

\subsection{Local versus non-local gravity}

 We can now use our general results in the previous sections, apply them to specific forms of $\cF(\square)$, and compare local and non-local quadratic curvature gravity. What we aim to show is that non-local gravity admits bouncing solutions with both positive matter energy density, and no instabilities (ghost/tachyonic). We have ensured that the energy density of each matter component remains non-negative. It is worth checking whether the gravitational part of the action remains ghost-free or not.
 
Note that both these criteria, that the gravity and matter sectors remaining ghost-free, may not be satisfied simultaneously in a local theory of gravity. Since our $\cosh$ bouncing solution (\ref{coshbounce}) is dynamical, we will look at its late time behavior instead for simplicity\footnote{Note that in \cite{Kumar:2020xsl}, the second order action for the non-local theory in (\ref{action1}) was derived around the background ansatz $\square R= r_1R+r_2$ given in (\ref{ansatz}) in the absence of any matter (but in the presence of $\Lambda$), with the tracelessness conditions in (\ref{condi}) and also $\cF_{1}=0$. This vacuum background solution corresponds to the last row in Table (\ref{table1}). In particular, for the dynamical FLRW cosh bouncing background given in (\ref{coshbounce}), it was found that at the linearized level, there exists a single scalar degree of freedom which can be made free from ghosts/tachyons and it has oscillatory and bounded evolution across the bounce. We leave the interesting task of similarly analyzing the perturbative spectrum around an anisotropic Bianchi-I background (\ref{B-Im}) for future investigations.}. At late times, the Ricci scalar for a cosh bouncing background goes to a constant given by $R=6r_{1}$ (see end of Sec.(\ref{sec2})). Since $r_1>0$, this corresponds to dS. The correct form of $\cF(\square)$ which avoids the introduction of propagating ghost or tachyonic degrees of freedom was already derived for maximally symmetric spacetimes in \cite{Biswas:2016egy}, and is given by
\be \label{fbox}
\cF(\square) = \frac{1 + \frac{24 r_{1}}{M_{P}^{2}} f_{0} - e^{\alpha(\square)}}{\frac{24 r_{1}}{M_{P}^{2}} \left( 1+ \frac{\square}{2r_{1}} \right)}
\ee
where $\alpha (\square)$ is some entire function of the $\square$ operator. This choice of $\cF(\square)$ ensures that the spin-2 quadratic action has no new zeros apart from the one present in the GR limit. Moreover, this choice introduces no new zeros in the spin-0 quadratic action, and so there is no Brans-Dicke type scalar. The above choice of ghost-free $\cF(\square)$ can satisfy the conditions required to ensure positive energy densities of all matter components driving a cosh bouncing universe (\ref{coshbounce}) in a Bianchi-I background (\ref{B-Im}). For concreteness, we assume the simplest choice of $\alpha(\square) = \square/M_{s}^{2}$.

 Let us consider the case where all the three matter components ($k-$matter, radiation and stiff matter) are present in the model. For $\rho_{k0}>0, \rho_{r0}>0$ and $\rho_{s0}>0$, we need to satisfy the conditions in (\ref{condi2}) (see also first row in Table (\ref{table1})). It is easy to verify that these conditions can be satisfied for a range of $(f_{0},r_{1},\sigma_{0})$ values. One such choice is, for instance $f_{0} = -1698.43, r_{1} = M_{s}^{2} = 0.01 M_{P}^{2}$ and $\sigma_{0} = 0.01 M_{P}$.


Now consider a fourth order local theory where the form factor $\cF(\square)$ is truncated at a finite order in derivatives, like $\cF(\square) = f_{0}$. Now, we can again satisfy the conditions for positivity of energy density in the sixth row in Table (\ref{table1}), for which $\cF_{1} = f_{0}<0, \cF'_{1}=0$ and $\cF_{2}=0$. Then, for a chosen value of $\sigma_{0}^{2}$, these can be satisfied for $0<r_{1}<\frac{5 \sigma_{0}^{2}}{3}$ and $\cF_{1} = \frac{3 r_{1}+\sigma_{0}^{2}}{36 r_{1}^2 - 60 r_{1} \sigma_{0}^{2}}$. However, $f_{0}<0$ signals the presence of a tachyonic instability \cite{Schmidt:2006jt}; we cannot simultaneously satisfy positivity of energy density and absence of tachyonic instability in this example local theory.

\section{Conclusions} \label{sec5}

In this paper, we have studied for the first time anisotropic but homogeneous Bianchi-I background solutions in non-local quadratic in Ricci scalar gravity in the presence of a cosmological constant, with a matter sector that turns out to include both perfect fluids and a fluid with anisotropic pressures. We show that this theory admits a bouncing scenario in the anisotropic background without violating positivity of energy density of each matter source. Unlike local higher derivative gravity, an anisotropic bounce in non-local gravity leads to late time FLRW (and eventually dS) without ghost/tachyonic modes in the late time dS phase. Spacetime anisotropies grow slower than in GR during the contraction phase, peak to a finite value at the bounce point, and then decrease during the expansion phase as the universe asymptotes to isotropic and homogeneous FLRW, and eventually dS. We found that driving a bounce in a Bianchi-I background necessarily requires anisotropic pressures in the $k$-matter sector, unless the form factor $\cF(\square)$ in the action obeys a certain constraint. One could now compute perturbations and check if they are free from any instabilities, especially around the dynamical bouncing background. Also, it would be interesting to see what the corresponding matter sector looks like for an exponential bounce in Bianchi-I spacetimes where $a (t) \sim e^{t^{2}}$, another known solution of the ansatz (\ref{ansatz}) used to simplify the EoM (\ref{EoM}). It is also important to check if there exists a non-singular cosmological evolution in non-local gravity in other Bianchi spacetimes. Furthermore, in this work, we have only studied the effect of having a non-local quadratic Ricci scalar term in the action, but it will be curious to see how bouncing scenarios evolve once we include non-local quadratic in Ricci and Weyl tensor terms too. We leave all these interesting questions for future investigations.

\section*{Acknowledgements}
We dedicate this work to the memory of John D. Barrow. We thank Tirthabir Biswas, Che-Yu Chen, Alexey S. Koshelev, Sohyun Park, Alexei A. Starobinsky and Paul J. Steinhardt for useful discussions. AM is supported by Netherlands Organization for Scientific Research (NWO) Grant No.680-91-119. JP is supported by the China Scholarship Council. KSK acknowledges  financial support  from  JSPS  and KAKENHI   Grant-in-Aid   for   Scientific   Research   No. JP20F20320. KSK was also supported partly by NWO Grant No.680-91-119.

\appendix
\section{Appendix: $k$-matter as a perfect fluid scalar field} \label{appendix}
It is interesting to consider the simplest case of $k$-matter being represented by a perfect fluid, canonical scalar field in a Bianchi-I background (\ref{B-Im}) with the Lagrangian $\Lc = - (\partial \phi)^{2}/2 - V(\phi)$. See Table (\ref{table1}) for a list of cases where $k$-matter can be described by a perfect fluid. We assume that it is described by a homogeneous and time-dependent scalar field $\phi(t)$. Using
\be
\ba
\rho &= \dot{\phi}^{2}/2 + V(\phi), \qquad p = \dot{\phi}^{2}/2 - V(\phi), \\
\rho_{k} &= \rho_{k0} a^{-2}, \qquad p_{k} = w_{k} \rho_{k}, \quad \text{ where }  w_{k} = -1/3, \\
H &= \sqrt{\frac{r_{1}}{2}} \tanh\left(  \sqrt{\frac{r_{1}}{2}} t \right), \qquad \text{for } a = \cosh \left(  \sqrt{\frac{r_{1}}{2}} t \right),
\ea
\ee
we can deduce an oscillatory potential given by
\be
V(\phi) = \frac{2 \rho_{k0}}{3} \cos^{2} \left( \sqrt{\frac{3 r_{1}}{4 \rho_{k0}}} \phi \right).
\ee
The solution for $\phi(t)$ is
\be
\phi(t) = \sqrt{\frac{4 \rho_{k0}}{3 r_{1}}} \ gd\left(\sqrt{\frac{r_{1}}{2}}t \right)
\ee
where $\phi(t)$ runs from $- \frac{\pi}{2} \sqrt{\frac{4 \rho_{k0}}{3 r_{1}}}$ at $t=-\infty$ to $\frac{\pi}{2} \sqrt{\frac{4 \rho_{k0}}{3 r_{1}}}$ at $t=\infty$.

%
%


\phantomsection
\addcontentsline{toc}{section}{References}
\providecommand{\href}[2]{#2}\begingroup\raggedright\endgroup

\end{document}